# Rural Healthcare Access and Supply Constraints: A Causal Analysis[*]

Vitor Melo [†]  Liam Sigaud [‡]  Elijah Neilson[§]  Markus Bjoerkheim[¶]

April 9, 2024

## Abstract

Certificate-of-need (CON) laws require that healthcare providers receive approval from a state board before offering additional services in a given community. Proponents of CON laws claim that these laws are needed to prevent the oversupply of healthcare services in urban areas and to increase access in rural areas, which are predominantly underserved. Yet, the policy could lower rural access if used by incumbents to limit entry from competitors. We explore the repeal of these regulations in five U.S. states to offer the first estimate of the causal effects of CON laws on rural and urban healthcare access. We find that repealing CON laws causes a substantial increase in hospitals in both rural and urban areas. We also find that the repeal leads to fewer beds and smaller hospitals on average, suggesting an increase in entry and competition in both rural and urban areas.

[*]We thank Matthew D. Mitchell, Ali Melad, and Andrew Blackburn for helpful comments and suggestions. [†]Clemson University and Mercatus Center at George Mason University. Email: vmelo@clemson.edu. [‡]Mercatus Center at George Mason University. Email: lsigaud@mercatus.gmu.edu.
[§]Southern Utah University - Dixie L. Leavitt School of Business. Email: elijahneilson@suu.edu.
[¶]Mercatus Center at George Mason University. Email: mbjoerkheim@mercatus.gmu.edu.

# 1 Introduction

Rural areas in the U.S. fare worse than their urban counterparts on many indicators of health, including mortality, chronic disease, and behavioral risk factors (Anderson et al., 2015). Moreover, the rural disadvantage in mortality and life expectancy has grown over recent decades (Singh and Siahpush, 2014; Snyder, 2016). A long-standing challenge to promoting health in rural areas is ensuring adequate access to healthcare services (Douthit et al., 2015; Iglehart, 2018). Persistent shortages of credentialed healthcare providers and long driving times to healthcare facilities make it difficult for many residents of sparsely populated areas to receive needed medical care. In recent years, access challenges have grown in many parts of the country. For example, 64 rural U.S. hospitals closed between 2013 and 2017, more than twice as many as in the previous five years (Cosgrove, 2018).

Currently, 35 states and Washington, D.C. have certificate-of-need (CON) laws. These laws require that healthcare providers receive approval from the government before offering additional services in a given community. In a state with a CON program, a government-controlled planning board makes decisions about whether the construction of a new healthcare facility or the expansion of an existing facility's services in a given geographic area is justified in light of community needs. A key objective of the National Health Planning and Resources Development Act of 1974, a federal law that heavily incentivized states to adopt CON regulations, was to improve rural access to healthcare (Mitchell, 2017; Smith and Pattni, 2020). In this paper, we offer the first estimate of the causal effects of CON laws on rural and urban hospital access. We find that repealing CON laws causes a substantial increase in hospitals in both rural and urban areas. We also find that the repeal leads to fewer hospital beds and smaller hospitals on average, suggesting an increase in entry and competition in both rural and urban areas.

Proponents of CON programs often claim the regulations ensure that care will be accessible to residents of geographically underserved, rural communities. A hospital network seeking to build a new hospital, for example, may expect to earn more profits by building a facility in an urban setting than in a rural area. In the absence of CON constraints, the facility is more likely to be built in an urban area. But CON laws allow state regulators to block the hospital's construction in an urban area, leading the hospital network to fall back to its second-best – but still potentially profitable – choice of constructing the facility in a rural area. The West Virginia Health Care Authority, the agency tasked with administering the state's CON program, argues: "In West Virginia, the CON program offers some protection for small, often financially fragile, rural hospitals and the underinsured population they serve by promoting the availability and accessibility of services" (West Virginia Health Care Authority, 2023). The Kentucky Hospital Association has claimed that repealing CON laws "would be another nail in the coffin for rural communities" (Kentucky Hospital Association, 2020). Thus, CON laws could improve rural access to care by allowing regulators to prevent the expansion of urban healthcare infrastructure, leaving rural markets as the next-best alternative for profit-seeking firms.

Yet, CON laws could lower rural access if used by incumbents to limit competition. In many states, the CON application process is lengthy and expensive. Larger and more influential healthcare entities may be better positioned to navigate the bureaucratic hurdles and exert political pressure to secure favorable outcomes (Schonbrun, 1978). Indeed, Stratmann and Monaghan (2017) find that political contributions affect the likelihood of CON application approval. Maureen Ohlhausen, former Commissioner of the U.S. Federal Trade Commission, asserts that CON laws "effectively serve primarily, if not solely, to assist incumbents in fending off competition from new entrants" (Ohlhausen, 2015). Courts have repeatedly struck down CON restrictions for being blatantly anti-competitive, such as when "a state program was found to have denied a proprietary hospital's application because of a hidden preference for existing facilities" (Simpson, 1985). In other words, the claim that CON laws are needed to promote rural healthcare access may be a ploy to conceal the real motive behind such policies: reducing competition.

We examine the effects of CON laws on rural and urban healthcare access by exploring the repeal of CON laws that covered hospital facilities in five U.S. states: North Dakota, Pennsylvania,

Nebraska, Ohio, and New Hampshire.[1] These states vary across many dimensions that affect the healthcare system, including population density, demographics, political orientation, and economic performance. Consequently, our results can be generalized to other states that may contemplate similar reforms. We apply a staggered difference-in-differences method proposed by Callaway and Sant'Anna (2021) to estimate the average treatment effect of repealing CON laws. We choose this method due to the potential biases associated with using a simple two-way fixed effects approach when there is heterogeneity in treatment timing (Goodman-Bacon, 2021).[2]

We find that the repeal of hospital CON laws causes an increase of 3.8% and 3.9% additional hospitals per capita in rural and urban areas, respectively. Moreover, the growth in the number of facilities begins between 3 and 6 years after the reform, consistent with the expected lag in opening new facilities. We also find that hospital beds per capita decline by 4.9% and 8% in rural and urban areas, respectively. Finally, we find that the average number of beds per hospital goes down by 5.1% and 14.8% in rural and urban areas, respectively, showing that the repeal caused treated states to have more but smaller hospitals. We find strong evidence in support of the common trends assumption required for the causal interpretation of our results. Our analysis undermines the claim that CON laws unequivocally increase rural healthcare access. Our results suggest that this claim may instead be an excuse to shield powerful incumbents from competition, since CON laws are associated with fewer but larger hospitals in both rural and urban areas.

Our work contributes to several strands of existing research. First, our work adds to a large literature on the effects of CON laws on the healthcare system. These laws were designed to prevent unnecessary proliferation of facilities and services, but their impact varies (Ho et al., 2009; Kim et al., 2016). Research indicates that CON laws can influence prices (Noether, 1988; Nyman, 1994; Bailey et al., 2016), utilization (Salkever and Bice, 1977), competitiveness (Paul et al., 2019),

expenditures (Conover and Sloan, 1998; Hellinger, 2009; Rahman et al., 2016; Bailey, 2019; Ettner

---

[1] We emphasize that we only consider the repeal of CON laws in states that covered hospital beds. Previous papers often consider the repeal of any CON law, regardless of whether that law covers hospitals, which is likely to lead to measurement errors and biased estimates. We, instead, compare states that repealed hospital CON laws during our study period to states that enforced hospital CON laws during the entire study period.

[2] The treated states in our sample repealed their hospital CON laws in different years: North Dakota (1995), Pennsylvania (1996), Nebraska (1997), Ohio (1998), and New Hampshire (2016).

et al., 2020; Melo and Neilson, 2022), service provision (Robinson et al., 2001; Cutler et al., 2010), health outcomes (Ho, 2006; Ho et al., 2009; Kim et al., 2016; Bailey, 2018; Fayissa et al., 2020; Myers and Sheehan, 2020; Chiu, 2021; Stratmann, 2022; Gaines and Cagle, 2023), and resiliency to the COVID-19 pandemic (Mitchell and Stratmann, 2021; Roy Choudhury et al., 2022).

Relatively little is known, however, about the causal effects of CON laws on rural healthcare access and delivery. D'aunno et al. (2000) study the market and institutional determinants of organizational change in rural hospitals, focusing specifically on the factors that make a rural hospital more likely to change the types of services it offers. The authors find that rural hospitals subject to stronger CON regulation are less likely to change. Stratmann and Baker (2020) explore the relationship between CON laws in rural states and measures of potentially avoidable healthcare utilization and spending. They find that residents of counties restricted by CON laws spend more per Medicare beneficiary and have higher utilization rates in ambulance services, emergency departments, and re-admissions than in counties without CON regulations. Herb et al. (2021) study the association between CON laws and travel time to radiation oncology facilities in rural and urban areas and find mixed results. Stratmann and Koopman (2016) find that the presence of CON laws is predictive of fewer hospitals and ambulatory surgical centers (ASCs) per capita. The authors note that "while we are able to present correlations, we do not have an identification strategy that would allow us to provide any causal interpretation to our results." Yet, estimating the causal effects of CON laws on rural health access is important to inform policy decisions regarding the repeal or implementation of these laws. We extend Stratmann and Koopman (2016)'s analysis by exploring the repeal of hospital CON laws in five U.S. states and deploying a robust causal inference method that accounts for the staggered timing of state reforms. Additionally, we examine a broader range of outcomes (i.e., total beds and average facility size) related to healthcare access that allow us to better understand how these laws shape the market for rural and urban hospital services.

Our work is important for three reasons. First, our findings imply that CON laws impose a substantial constraint on the supply of hospital services in both rural and urban areas. Second, our results indicate that CON laws stifle competition in the hospital sector by preventing smaller facilities from entering the market. This finding is consistent with the notion that CON boards often exhibit regulatory capture by large incumbent providers. Third, our results are at odds with a popular justification for CON laws – namely, that such restrictions are used to divert investments in healthcare infrastructure from urban areas to rural communities.

## 2   Data

We assemble 1991-2019 data on state CON laws from the American Health Planning Association and the National Conference of State Legislatures. Since our analysis focuses on access to hospital care, we define our treatment variable based on the presence or absence of CON restrictions on hospital facilities.[3] Figure 1 shows the status of hospital CON laws in the U.S. during our study period. Five states, colored in black, repealed their hospital CON laws during our study period; these constitute our treatment group. Twenty-nine states and Washington, D.C., colored in dark gray, enforced hospital CON laws throughout our study period; these constitute our control group. Sixteen states, colored in light gray, did not have hospital CON laws at any time during our sample period; we drop these states from our analysis.

We collect facility-level data from the Provider of Services (POS) files from the Centers for Medicare & Medicaid Services (CMS), which provide information on the universe of CMS-certified non-laboratory institutional providers in the country. The file is updated annually to reflect closures, mergers, and other changes to facilities' operational status. We construct a panel dataset of hospital facilities from 1991 to 2019. Given our interest in patient access to hospital services, we exclude hospitals that are not listed as active, hospitals that report zero beds, and hospitals with missing data for bed count.

To examine the heterogeneous effects of CON laws on rural and urban areas, we link the POS

---

[3]Our data on CON laws does not have sufficient granularity to detect subtle changes in CON laws (e.g., a state amending its CON program to exclude rural facilities). To our knowledge, no publicly available dataset exists with this type of information covering all 50 states for our sample period.

data to county-level rural-urban continuum codes developed by the U.S. Department of Agriculture (USDA).[4] The USDA codes place each county on a rural-urban spectrum, ranging from "central counties of metro areas of 1 million population or more" to "completely rural or less than 2,500 urban population, not adjacent to a metro area." For the purposes of our analysis, we combine multiple code categories into a dichotomous variable: inside a Metropolitan Statistical Area (MSA) or outside an MSA. An MSA is an economically-integrated agglomeration of one or more counties that contain a city of 50,000 or more inhabitants. Facilities located within an MSA in 2003 (roughly at the mid-point of our study period) are considered urban; facilities located outside an MSA in 2003 are considered rural.[5]

We derive three outcomes from the POS data, calculated separately for rural and urban populations of each state: (1) the number of hospital facilities per 100,000 residents, (2) the number of hospital beds per 100,000 residents, and (3) the average number of beds per hospital facility. The calculation for outcome (1) is given by

$$\text{Number of Hospital Facilities per 100,000}_{t,s,r} = \frac{\text{Number of Facilities}_{t,s,r}}{\text{Population}_{t,s,r}} * 100,000, \quad (1)$$

where t indexes time (measured in years), s indexes states (including Washington, D.C.), and r indexes rural/urban areas. Number of Facilities is the number of hospital facilities in operation in a given year, state, and rural or urban area. Population is the total non-institutionalized resident population. The ratio of Number of Facilities to Population is scaled by 100,000 to ease interpretation but has no effect on our underlying results. Outcome (2), the number of beds per 100,000

---

[4]In rare cases healthcare facilities could not be matched to county identifiers and were dropped from the analysis; this affected about 1% of the facilities in the raw POS dataset. Typically, the county FIPS codes for these providers were missing, did not exist, or reflected old county identifiers that had been discontinued when county lines were redrawn.

[5]A small minority of counties changed MSA status during our sample period; to maintain a consistent unit of analysis, we ignore these re-classifications. In other words, if a county was classified as rural (urban) in 2003, we maintain that classification throughout our study period (1991-2019), regardless of whether in any other year the county was re-classified as urban (rural). Moreover, three jurisdictions – Washington, D.C., New Jersey, and Rhode Island – did not contain any non-MSA counties in 2003. As a result, our rural dataset has 87 fewer observations (3 states × 29 years) than our urban dataset.

Figure 1: Hospital Certificate-of-Need Regulation in the United States

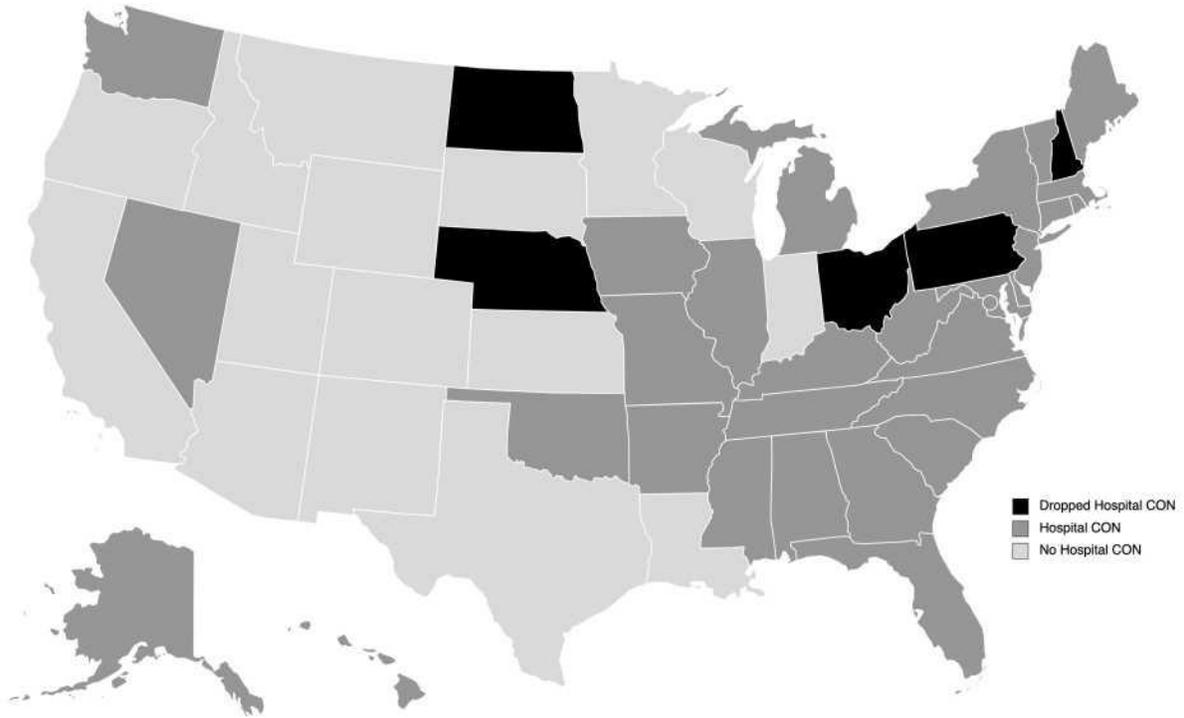

Notes: This map shows which states dropped hospital CON laws during our sample period (black), and which states did (dark gray) and did not (light gray) have hospital CON laws throughout our sample period. Treated states consist of those states that dropped hospital CON laws during our sample period. All states that had hospital CON laws throughout our sample period are considered controls. States that did not have hospital CON laws at any point during our sample period are not used in our analysis. Data source: American Health Planning Association (AHPA) and the National Conference of State Legislatures (NCSL).

residents, is calculated analogously to Equation 1.

To illustrate these computations: we determine the number of rural hospitals in Wyoming in 1991 per 100,000 residents by dividing the total number of hospitals in rural counties of Wyoming in 1991 by the total resident population of rural counties in Wyoming in 1991 and multiplying the quotient by 100,000. We perform the same calculation for other states, years, and rural/urban areas.

Our third outcome, average beds per facility, is simply the ratio of the total number of hospital beds in a given state, year, and rural/urban area to the total number of hospital facilities in that state, year, and rural/urban area. Formally, this computation can be written as

Table 1: Descriptive Statistics

| | Treated States | | | Untreated States | | |
|---|---|---|---|---|---|---|
| Variable | Mean | SD | Median | Mean | SD | Median |
| Panel A. Rural Areas | | | | | | |
| Hospital Beds/100k | 400.19 | 93.32 | 389.43 | 336.96 | 137.44 | 328.57 |
| Hospital Facilities/100k | 6.22 | 3.69 | 3.86 | 4.16 | 1.55 | 4.03 |
| Beds Per Facility | 84.57 | 37.94 | 86.58 | 87.74 | 32.16 | 89.19 |
| Real Income Per Capita | 38,649.36 | 7,972.73 | 36,721.05 | 35,993.90 | 8,924.80 | 33,628.5 |
| Proportion Aged 65+ | 0.17 | 0.02 | 0.17 | 0.16 | 0.03 | 0.16 |
| N | | 145 | | | 783 | |
| Panel B. Urban Areas | | | | | | |
| Hospital Beds/100k | 487.66 | 131.25 | 482.74 | 475.64 | 193.94 | 435.01 |
| Hospital Facilities/100k | 2.39 | 0.66 | 2.15 | 2.08 | 0.78 | 1.87 |
| Beds Per Facility | 208.35 | 47.81 | 199.67 | 237.76 | 62.18 | 233.14 |
| Real Income Per Capita | 45,270.51 | 6,505.35 | 45,237.93 | 45,058.18 | 8,669.40 | 43,698.7 |
| Proportion Aged 65+ | 0.13 | 0.02 | 0.13 | 0.13 | 0.02 | 0.13 |
| N | | 145 | | | 870 | |

Notes: This table reports the mean, standard deviation, and median of each of our outcome variables and covariates separately for treated states (those that dropped hospital CON laws during the sample period) and untreated states (those that had hospital CON laws throughout the sample period). Three jurisdictions – Washington, D.C., New Jersey, and Rhode Island – do not contain any non-MSA counties. As a result, Panel A (rural areas) has 87 fewer observations (3 states × 29 years) than Panel B (urban areas). Data sources: 19912019 Centers for Medicare and Medicaid Services' (CMS) Provider of Services files, U.S. Bureau of Economic Analysis, and National Cancer Institute SEER Program.

$$\text{Beds per Facility}_{t,s,r} = \frac{\text{Total Number of Beds}_{t,s,r}}{\text{Total Number of Facilities}_{t,s,r}} \qquad (2)$$

where t indexes time (measured in years), s indexes states (including Washington, D.C.), and r indexes rural/urban areas.

We report descriptive statistics for all outcomes in Table 1 separately by rural/urban status for treated and untreated states.

## 3 Empirical Strategy

To identify the causal effect of the repeal of CON laws on hospital access, we leverage the staggered repeal of CON laws that covered hospital beds in North Dakota (1995), Pennsylvania (1996), Nebraska (1997), Ohio (1998), and New Hampshire (2016). While traditional two-way fixed effects (TWFE) difference-in-differences regressions have long served as the primary models for staggered adoption research designs, recent studies have demonstrated their limitations in estimating the average treatment effect on the treated, particularly when treatment adoption is staggered over time and when average treatment effects vary across groups and over time (De Chaisemartin and d'Haultfoeuille, 2020; Borusyak et al., 2021; Callaway and Sant'Anna, 2021; Goodman- Bacon, 2021; Imai and Kim, 2021; Sun and Abraham, 2021; Athey and Imbens, 2022). We investigate the potential bias that would be present in our setting by implementing the decomposition- method proposed by Goodman-Bacon (2021), and present this along with traditional two-way fixed effects estimates in Appendix A1.[6]

We proceed by implementing the difference-in-differences estimator described in Callaway and Sant'Anna (2021) as our main analysis. This methods allows for estimation and inference on interpretable causal parameters even in the presence of arbitrary treatment effect heterogeneity and dynamic effects, thereby circumventing the issues of interpreting results of standard two-way fixed effects regressions as causal effects.

Callaway and Sant'Anna (2021) emphasize the identification of what they call the "group-time average treatment effect" (AT T (g,t)), which is the average treatment effect for group g at time t, where a "group" is defined by the time period when the units are first exposed to the treatment.[7] More specifically, the AT T (g,t)'s are defined as the difference in outcomes between the most recent period before repealing hospital CON and the current period t for a particular group g, adjusted by the analogous path of outcomes experienced by a comparison group.[8] This latter path is intended to represent the path of outcomes that the state in group g would have experienced had they not repealed hospital CON. Thus, if there is no treatment anticipation behavior and the parallel trends assumption holds, the ATT(g,t) identifies the causal effect of repealing hospital CON for a group g at time t .

---

[6]We find the influence of the problematic Late vs Early treatment-group comparisons to be small, with weights ranging from 2.8% in the case of our Rural outcomes to 3.1% for our Urban outcomes. The point estimates from TWFE are therefore comparable to those found in our main analysis, but less precise. See Appendix A1 for more details.

[7]In our context, each group consists of a single state since there are no cases in which multiple states repealed hospital CON laws in the same year.

[8]For our comparison group, we use only the states that had and maintained hospital CON regulations over the sample period (i.e. "never-treated" states).

One appealing aspect of the ATT (g,t) parameters is that they do not directly restrict heterogeneity with respect to the period in which units are first treated, or the evolution of treatment effects over time. Consequently, they can be directly used for learning about treatment effect heterogeneity, and to construct more aggregated causal parameters. We consider two such aggregations. First, to summarize the overall average effect of repealing hospital CON, we compute the average effect for each group (averaging the ATT (g,t)'s for each group across all post-treatment years) and then average these effects together across groups.[9] Thus, this aggregation captures the average effect of participating in the treatment experienced by all units that ever participated in the treatment. Table 2 reports these aggregated measures for each outcome variable within rural and urban areas. Second, we average ATT (g,t)'s according to event time – time relative to the repeal of hospital CON. This aggregation highlights treatment effect dynamics with respect to length of exposure to the treatment, and allows us to assess parallel pre-trends. Figure 2 plots these estimates separately for rural and urban areas with corresponding 95% confidence intervals for each outcome variable.

# 4  Results

## 4.1 Main Results

Table 2 presents our estimates of the average treatment effects on treated states (ATT) of removing hospital CON laws on the number of hospital facilities and hospital beds, both normalized by the resident population (expressed per 100,000 residents), as well as the number of beds per facility. We present the results separately by rural/urban status, and include robust standard errors clustered at the state level, as well as the baseline mean of each outcome variable within the treated states across all years prior to repealing hospital CON laws for comparison.

We find that removing CON laws leads to an increase in the number of operating hospitals in

---

[9]Alternatively, one could just average all of the identified ATT (g,t)'s together. As Callaway and Sant'Anna (2021) point out, however, this approach systematically puts more weight on groups that participate in the treatment longer.

Table 2: Effects of Dropping Hospital CON Laws on Access to Hospital Services

| | Facilities Per 100k | | Beds Per 100k | | Beds Per Facility | |
|---|---|---|---|---|---|---|
| | Rural | Urban | Rural | Urban | Rural | Urban |
| | (1) | (2) | (3) | (4) | (5) | (6) |
| Average Effect | 0.19*** | 0.09** | -20.81** | -37.87*** | -4.91*** | -29.70*** |
| Standard Error | (0.07) | (0.05) | (8.38) | (8.95) | (1.58) | (3.31) |
| Pre-Intervention Mean | 5.29 | 2.32 | 432.05 | 471.20 | 95.81 | 201.64 |
| N | 928 | 1,015 | 928 | 1,015 | 928 | 1,015 |

Notes: This table reports the average effects of dropping hospital CON laws on the number of hospital facilities per 100,000, beds per 100,000, and beds per facility within rural and urban areas. Robust standard errors, clustered at the state level, are in parenthesis, and significance levels are reported as *** = p<0.01, ** = p<0.05, * = p<0.1. The baseline mean of each outcome variable within the treated states across all years prior to dropping hospital CON laws is also reported. Three jurisdictions – Washington, D.C., New Jersey, and Rhode Island – do not contain any non-MSA counties. As a result, rural areas have 87 fewer observations (3 states × 29 years) than urban areas. Data source: 1991-2019 Centers for Medicare and Medicaid Services' (CMS) Provider of Services files.

both rural and urban areas. Our estimates tend to be of similar magnitude when expressed relative to the rural/urban mean in treated states. Specifically, in column (1) we estimate the causal effect of CON law repeal to be an increase of 0.19 hospital facilities per 100,000 residents in rural areas. This estimate is significant at the 1% level and the effect size is 3.6% of the pre-treatment mean (5.29) in treated states. Similarly, in column (2) we estimate the effect of CON law repeal in urban areas to be an increase of 0.09 facilities per 100,000 residents. This effect is significant at the 5% level and is approximately 3.9% of the urban mean (2.32) in treated states.

Although we estimate that CON law repeal leads to more facilities per resident, we also find a reduction in the number of beds per resident in both rural and urban areas. In column (3), we estimate the average effect to be a reduction of 20.81 beds per 100,000 residents in rural areas. This effect is significant at the 5% level and the effect size is approximately 4.8% relative to the rural pre-treatment mean (432.05). In column (4), we find the effect to be a reduction of 37.87 beds per 100,000 residents in urban areas. This effect is significant at the 1% level and is approximately 8% relative to the urban pre-treatment mean (471.2) in treated states. Mechanically, the result of fewer beds can only be due to adjustments on the intensive margin or extensive margins. Jointly, our findings of more facilities but fewer beds imply that the net effect of hospitals shrinking or closing outweighs the additional beds from new facilities entering the market.

The effects on facilities and beds can also be expressed in terms of the effect of repealing CON laws on the average size of hospital facilities. We find a reduction of 4.91 beds per facility in rural areas (column 5) and a reduction of 29.7 beds per facility in urban areas (column 6). These effects are both significant at the 1% level and translate to reductions of 5.1% (rural) and 14.7% (urban) relative to the mean in treated states.

The causal interpretation of all coefficients shown in Table 2 depends on the assumption that in a counterfactual where treated states were not treated, they would have followed the same trends as states in the control group. To assess the plausibility of this assumption, we follow common practice and inspect event study plots (Figure 2) that trace the effects of dropping hospital CON laws over time for each of our outcomes. We present coefficients for 10 pre-treatment years and 20 post-treatment years. Figure 2 shows that treated and control states followed very similar, statistically indistinguishable trends prior to treatment. This supports the causal interpretation of the estimates presented in Table 2.

The event study plots also provide insight into the dynamic effects of repealing CON laws across time. Panel (a) of Figure 2 shows a clear uptick in the number of facilities in both rural and urban areas starting 3-6 years after CON repeal. This lag between repeal and entry of additional facilities is plausible given the time needed to raise funds and complete the physical construction of a new hospital facility. Panel (b) of Figure 2, which focuses on the number of beds, shows different patterns in rural and urban areas. In rural areas, we find a gradual reduction in beds starting about 3 years after repeal; these effects appear to grow over our sample period. On the other hand, the estimated reduction in urban areas is immediate and appears to stay relatively constant. Finally, panel (c) of Figure 2 shows the dynamic effects of repealing CON laws on the average facility size. We find a slow gradual decline in rural areas and larger, steadily-growing reductions in urban areas.

Based on these findings, we conclude that the repeal on CON led to an increase in the number of facilities but a decrease in the average size of these facilities. The increase in the number of

Figure 2: Event Study Plots of the Effects of Dropping Hospital CON Laws on Access to Hospital Services

(a) Hospital Facilities

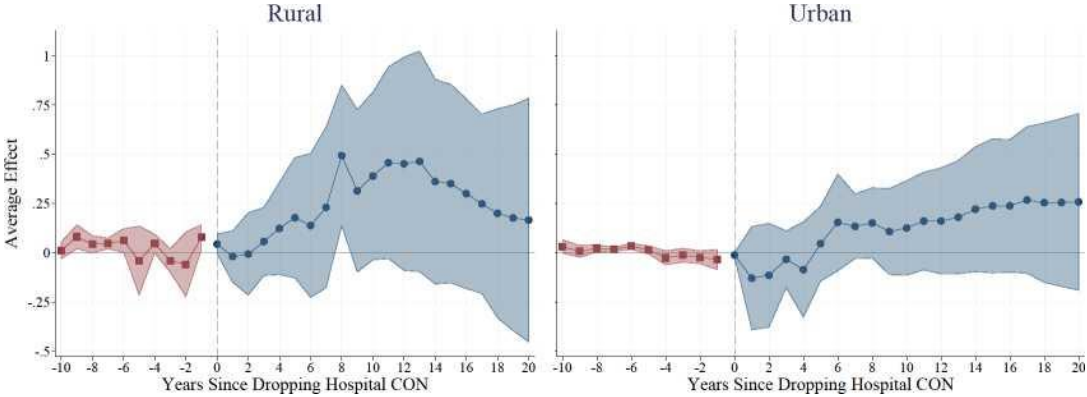

(b) Hospital Beds

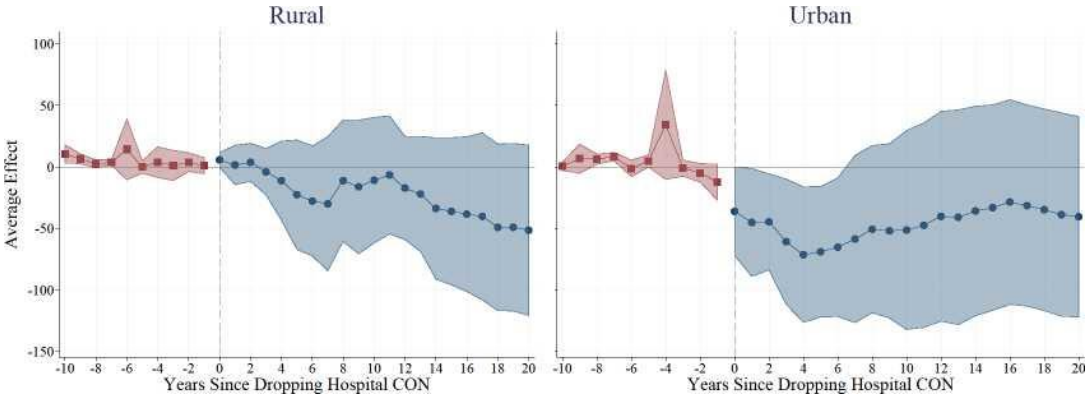

(c) Hospital Beds Per Facility

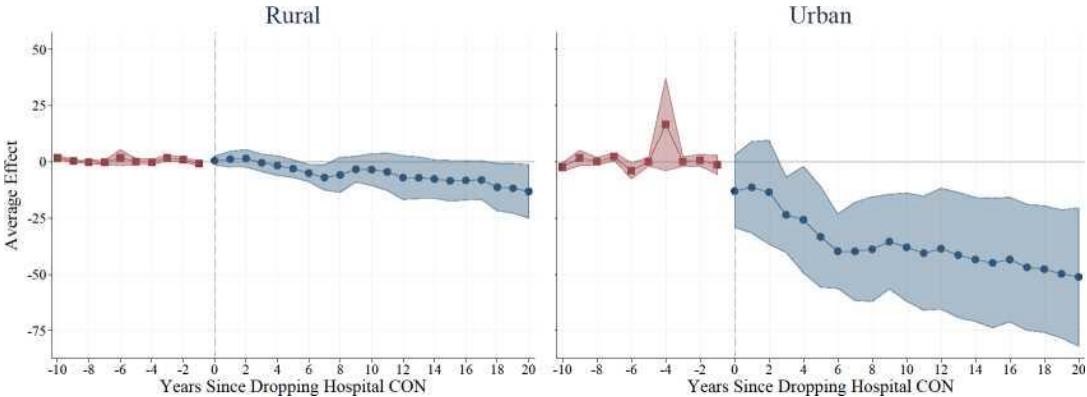

Notes: This figure shows event study plots of the average effects of dropping hospital CON laws on the number of hospital facilities per 100,000, the number of hospital beds per 100,000, and the number of hospital beds per facility within rural and urban areas. The shaded regions represent 95% confidence intervals for each estimate. Data source: 1991-2019 Centers for Medicare and Medicaid Services' (CMS) Provider of Services files.

facilities we document is likely to make basic medical services more accessible, particularly in rural communities where long driving times represent a major barrier to care (Arcury et al., 2005; Syed et al., 2013). However, the accompanying reduction in hospital beds may indicate that larger facilities are trimming specialty services (e.g., maternity wards or psychiatric units) that are not being replaced by new entrants. This could undermine access to care for certain populations. Moreover, the reduction in hospital beds may make the health system less resilient to sudden crises like the COVID-19 pandemic.

## 4.2 Robustness Analysis

To test the robustness of our main analysis we rerun the analysis from Table 2 while including real per capita income and the share of the population aged 65 or older as matching arguments. Real per capita income, which we obtain from the U.S. Bureau of Economic Analysis, helps to account for differences between states in the health of the economy. We calculate the share of the population aged 65 or older based on age-specific demographic data drawn from National Cancer Institute's Surveillance, Epidemiology, and End Results (SEER) Program. The Doubly Robust Inverse Probability Weighting (DR-IPW) we implement will then produce a consistent estimate of the treatment effect if either an outcome-based regression model or propensity scores based on the covariates, is a correctly specified model (Sant'Anna and Zhao, 2020; Callaway and Sant'Anna, 2021). We are unable to include Hawaii in this analysis because age-specific population estimates for rural and urban areas of Hawaii are not available prior to the year 2000.

    The results are presented in Table 3 and are qualitatively very similar to those in our main analysis. The point estimate for facilities in rural areas is identical and remains significant at the 5% level. The point estimate for facilities in Urban areas is larger in the alternative specification and is now significant at the 1% level, as opposed to the 5% level. The point estimates for beds and beds per facility all shrink slightly in Rural areas and grow slightly in Urban areas, but remain statistically significant at the same levels as in the main analysis. We interpret these results as indicating our main analysis is robust to alternative specifications and is not overly reliant on the

Table 3: Effects of Dropping Hospital CON Laws on Access to Hospital Services (DR-IPW)

|  | Facilities Per 100k | | Beds Per 100k | | Beds Per Facility | |
|---|---|---|---|---|---|---|
|  | Rural (1) | Urban (2) | Rural (3) | Urban (4) | Rural (5) | Urban (6) |
| Average Effect | 0.19** | 0.15*** | -11.26** | -49.25*** | -3.12*** | -37.43*** |
| Standard Error | (0.09) | (0.03) | (5.16) | (6.33) | (1.08) | (3.01) |
| N | 899 | 986 | 899 | 986 | 899 | 986 |

Notes: This table reports the average effects of dropping hospital CON laws on the number of hospital facilities per 100,000, beds per 100,000, and beds per facility within rural and urban areas using the doubly robust inverse probability weighting (Callaway and Sant'Anna, 2021) estimator. Robust standard errors, clustered at the state level, are in parenthesis, and significance levels are reported as *** = p < 0.01, ** = p <0.05, * = p < 0.1. Covariates used include real income per capita and the proportion of the population aged 65 or older. Data sources: 1991-2019 Centers for Medicare and Medicaid Services' (CMS) Provider of Services files, U.S. Bureau of Economic Analysis, and National Cancer Institute SEER Program.

outcome-based approach.

We also report the results of a two-way fixed effects model in Table A1. We show the event study plots associated with each estimate from Table A1 in Figures A1 and A2 in the appendix. The results are qualitatively consistent with our main results from subsection 4.1. Yet, we do not rely our inference on the results from the two-way fixed effects model because of the well-known biases associated with this model in settings with heterogeneous treatment timing like ours. Moreover, some of our two-way fixed effects models likely violate the parallel trends assumption needed to interpret the results as causal. Although the bias from the two-way fixed effects model may be modest, we rely only on the results from Tables 2 and 3 since these results do not suffer from any biases caused by treatment timing heterogeneity.

Figure 3: Event Study Plots of the Effects of Dropping Hospital CON Laws on Access to Hospital Services (DR-IPW)

(a) Hospital Facilities

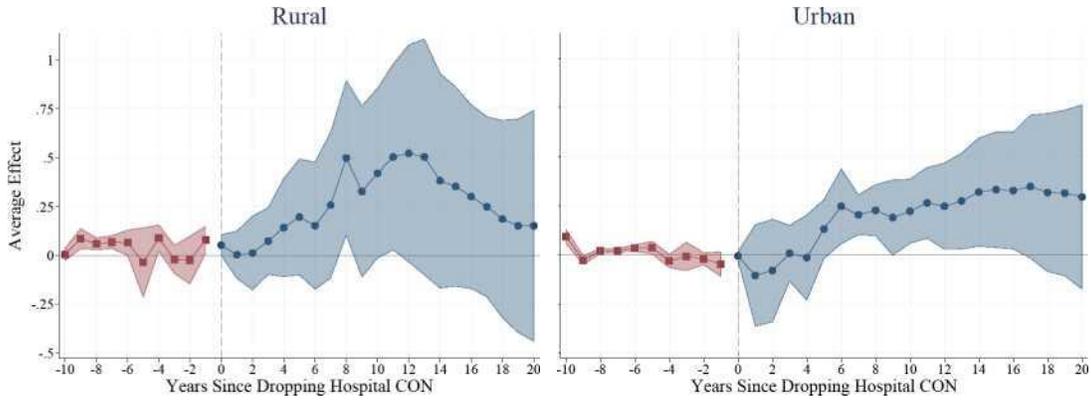

(b) Hospital Beds

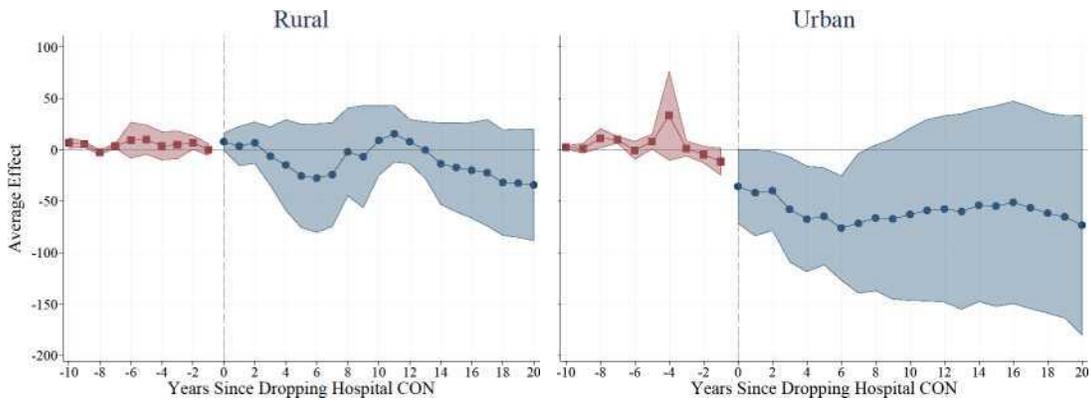

(c) Hospital Beds Per Facility

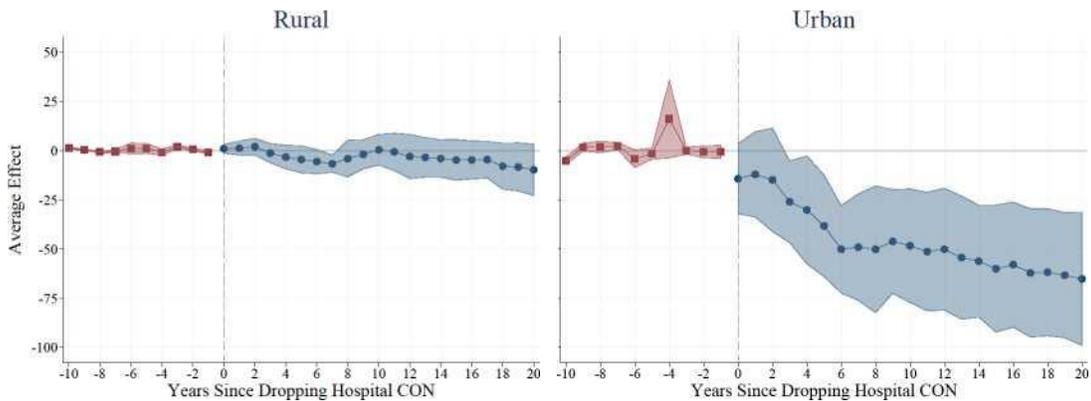

Notes: This figure shows event study plots of the average effects of dropping hospital CON laws on the number of hospital facilities per 100,000, the number of hospital beds per 100,000, and the number of hospital beds per facility within rural and urban areas using the doubly robust inverse probability weighting estimator. The shaded regions represent 95% confidence intervals for each estimate. Covariates used include real income per capita and the proportion of the population aged 65 or older. Data sources: 1991-2019 Centers for Medicare and Medicaid Services' (CMS) Provider of Services files, U.S. Bureau of Economic Analysis, and National Cancer Institute SEER Program.

# 5 Conclusion

The scarcity of healthcare services in rural areas of the U.S. remains a pressing public health concern. Over the past several decades, many rural communities have experienced an erosion of health services driven by hospital closures and a shrinking workforce. These trends may have helped fuel the widening gap in mortality and life expectancy between urban and rural areas of the U.S. CON laws are among the most important state regulatory constraints on the supply of healthcare services. Yet little is known about how these policies influence rural healthcare access.

Whether CON laws serve to expand or restrict rural healthcare access largely depends on how the laws are administered. It is possible, as proponents of CON laws often claim, that these regulations enable states to increase investments in rural healthcare services by denying providers the opportunity to expand in urban markets. Alternatively, CON laws may be exploited by incumbent providers to suppress any new healthcare investment and thereby reduce healthcare access generally, including in rural areas.

In this paper, we provide the first rigorous causal estimates of the effects of repealing CON laws on the hospital sector. We find that lifting these regulations causes a large, sustained increase in the number of hospital facilities per capita in both urban and rural areas and reduces the number of hospital beds per capita in both urban and rural areas. As a result, we find that CON laws cause a substantial decline in the average size of facilities, as measured by the number of beds per facility. Together, these results suggest that CON laws deter market entry by smaller hospitals. They also imply that repealing CON laws causes larger incumbent hospitals to shrink or close. Overall, our findings provide little support for the claim that eliminating CON laws jeopardizes rural healthcare access. Instead, our results suggest that CON laws may be used by existing providers to limit competition.

Our work suggests several avenues for future research. It is unclear from our results whether the decline in average facility size is caused by relatively large hospitals merely shrinking or closing altogether. This distinction has important implications for access to care, particularly for basic services. A related question – stemming from our finding that repealing CON laws results in fewer population-adjusted hospital beds – is whether facilities are reducing general care beds, paring back on specialty services, or a combination of both. Exploring these issues is needed to develop a full understanding of the impact of CON law repeal on access to care, and to inform effective policymaking.

# A Appendix

Table A1: Effects of Dropping Hospital CON Laws on Access to Hospital Services (Two-Way Fixed Effects)

|  | Facilities Per 100k | | Beds Per 100k | | Beds Per Facility | |
|---|---|---|---|---|---|---|
|  | Rural | Urban | Rural | Urban | Rural | Urban |
|  | (1) | (2) | (3) | (4) | (5) | (6) |
| **Panel A: Two-Way Fixed Effects** | | | | | | |
| Average Effect | 0.25 | 0.09 | -8.80 | -33.01 | -3.95 | -28.50*** |
| Standard Error | (0.16) | (0.12) | (21.65) | (30.81) | (3.00) | (9.98) |
| N | 928 | 1,015 | 928 | 1,015 | 928 | 1,015 |
| **Panel B: Two-Way Fixed Effects (Including Covariates)** | | | | | | |
| Average Effect | 0.29* | 0.12 | 6.96 | -31.66 | -2.23 | -30.42*** |
| Standard Error | (0.16) | (0.10) | (15.41) | (24.47) | (3.14) | (8.23) |
| N | 899 | 986 | 899 | 986 | 899 | 986 |

Notes: This table reports the average effects of dropping hospital CON laws on the number of hospital facilities per 100,000, beds per 100,000, and beds per facility within rural and urban areas using the traditional two-way-fixed effects estimators. Robust standard errors, clustered at the state level, are in parenthesis, and significance levels are reported as *** = p<0.01, ** = p<0.05, * = p<0.1. Covariates used in Panel B include real income per capita and the proportion of the population aged 65 or older. Data sources: 1991-2019 Centers for Medicare and Medicaid Services' (CMS) Provider of Services files, U.S. Bureau of Economic Analysis, and National Cancer Institute SEER Program.

Figure A1: Event Study Plots of the Effects of Dropping Hospital CON Laws on Access to Hospital Services (Two-Way Fixed Effects)

(a) Hospital Facilities

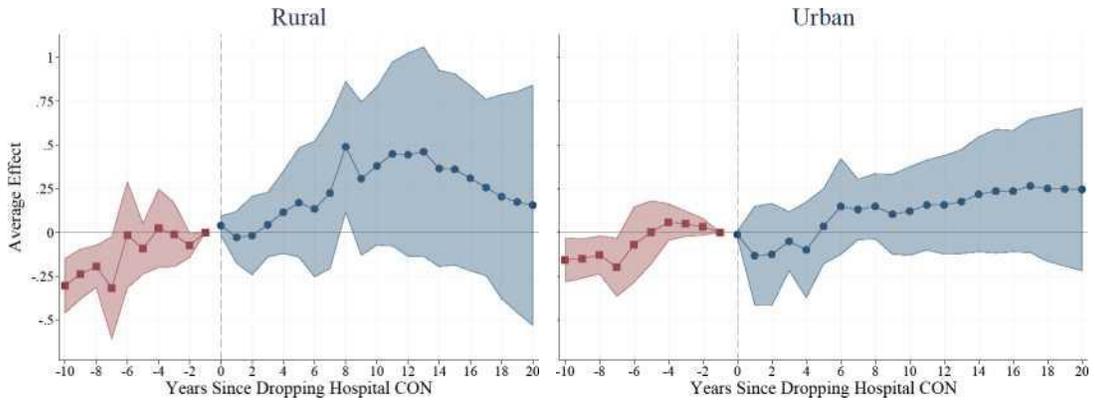

(b) Hospital Beds

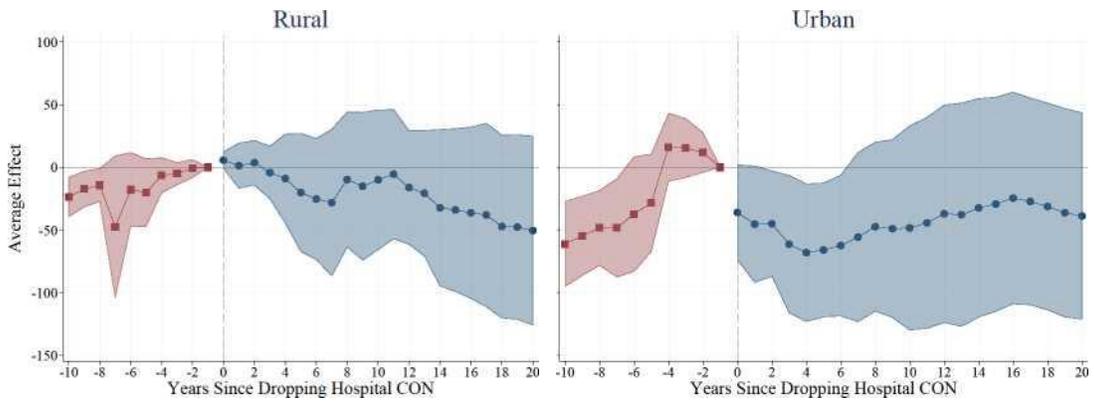

(c) Hospital Beds Per Facility

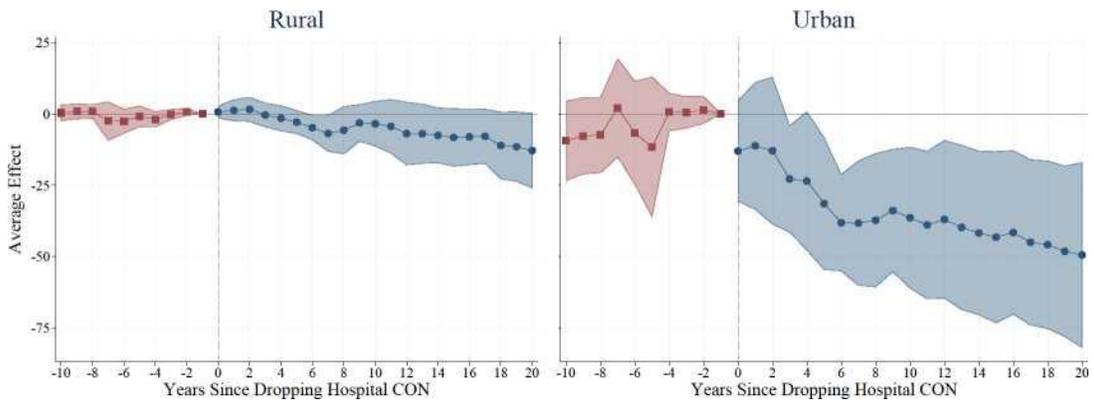

Notes: This figure shows event study plots of the average effects of dropping hospital CON laws on the number of hospital facilities per 100,000, the number of hospital beds per 100,000, and the number of hospital beds per facility within rural and urban areas using the traditional two-way fixed effects estimator. The shaded regions represent 95% confidence intervals for each estimate. Data source: 1991-2019 Centers for Medicare and Medicaid Services' (CMS) Provider of Services files.

Figure A2: Event Study Plots of the Effects of Dropping Hospital CON Laws on Access to Hospital Services (Two-Way Fixed Effects Including Covariates)

(a) Hospital Facilities

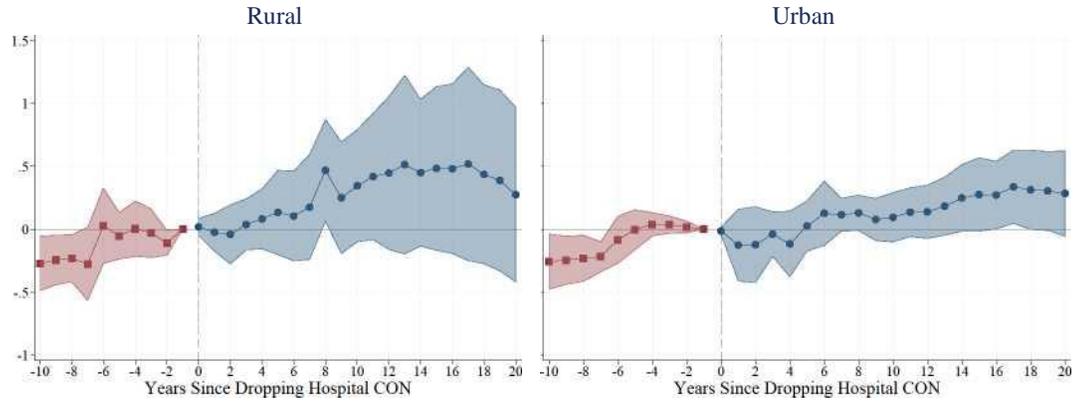

(b) Hospital Beds

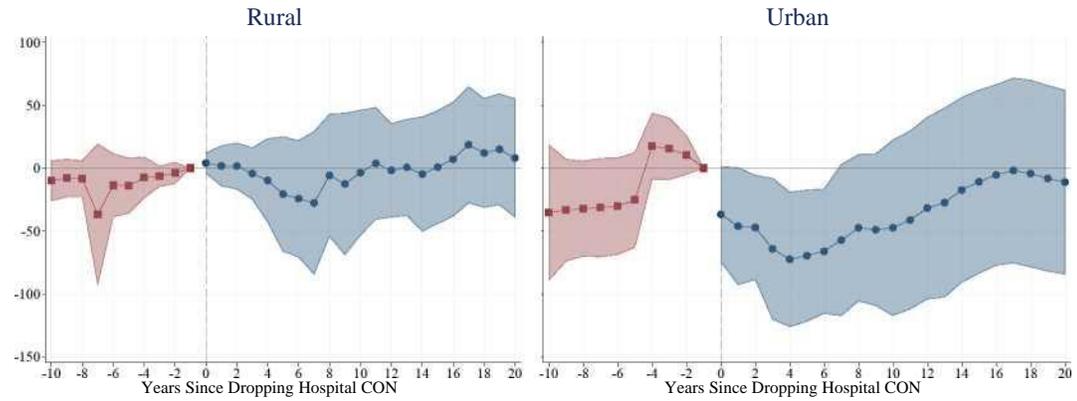

(c) Hospital Beds Per Facility

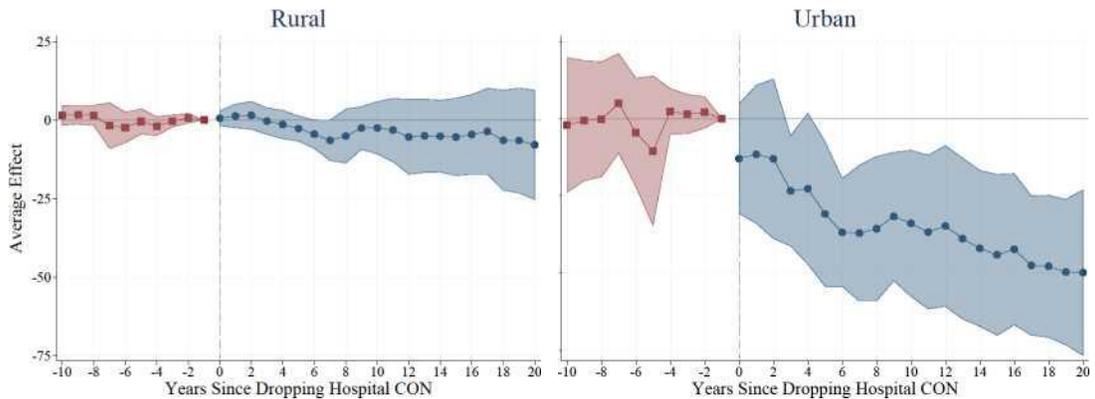

Notes: This figure shows event study plots of the average effects of dropping hospital CON laws on the number of hospital facilities per 100,000, the number of hospital beds per 100,000, and the number of hospital beds per facility within rural and urban areas using the traditional two-way fixed effects estimator. Covariates used in the estimation include real income per capita and the proportion of the population aged 65 or older. The shaded regions represent 95% confidence intervals for each estimate. Data source: 1991-2019 Centers for Medicare and Medicaid Services' (CMS) Provider of Services files, U.S. Bureau of Economic Analysis, and National Cancer Institute SEER Program.

Figure A3: Bacon Decomposition of Two-Way Fixed Effects Estimates

(a) Hospital Facilities

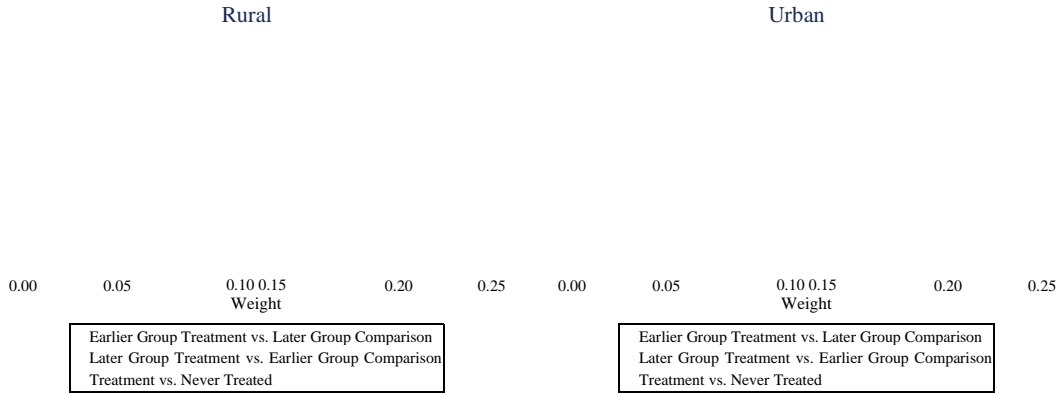

(b) Hospital Beds

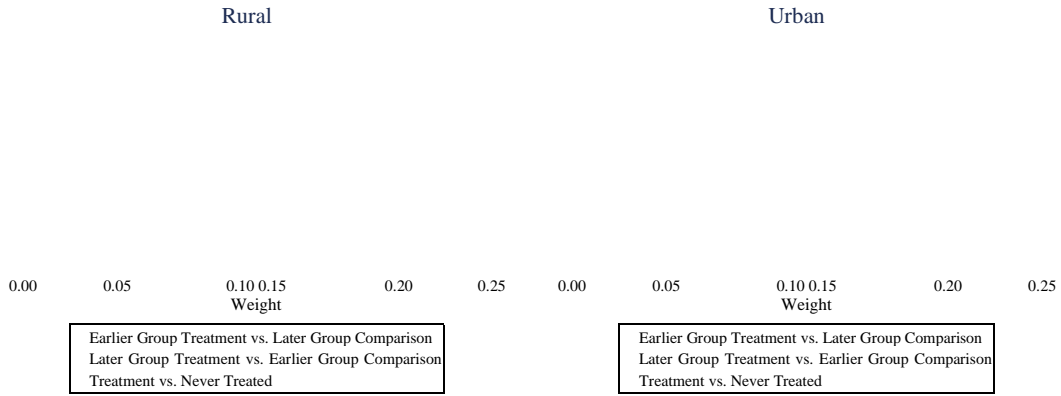

(c) Hospital Beds Per Facility

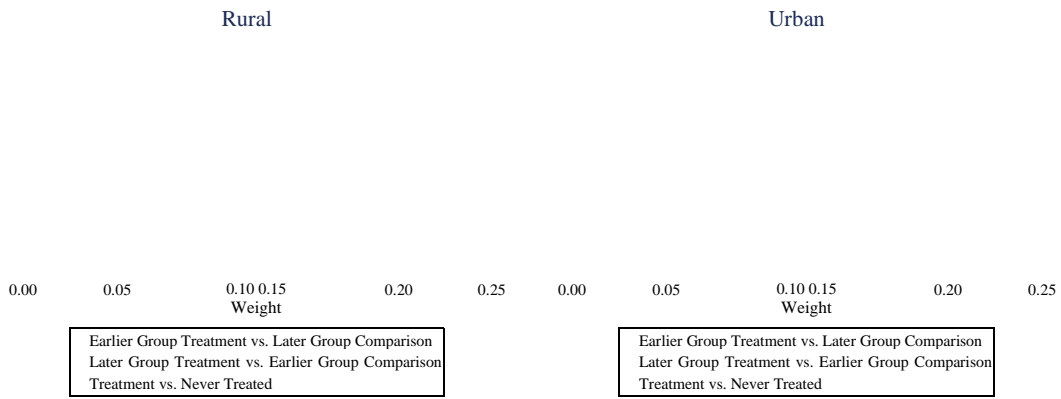

Notes: This figure shows the Bacon decomposition of our two-way fixed effects estimates in Panel A of Table A1. The estimated overall average treatment effect, denoted by the horizontal line in each plot, equals a weighted average of all possible two-group/two-period difference-in-differences estimates (Goodman-Bacon, 2021). The points on each plot represent each of these two-group/two-period estimates with their corresponding weights. Data source: 1991-2019 Centers for Medicare and Medicaid Services' (CMS) Provider of Services files, U.S. Bureau of Economic Analysis, and National Cancer Institute SEER Program.